\documentclass[preprint,12pt]{elsarticle}

\usepackage{amssymb}
\journal{the arXiv.org}

\begin{document}

\begin{frontmatter}

\title{Adaptive Control of Enterprise }

\author{Yuriy Ostapov}
\address{Institute of Cybernetics of NAS of Ukraine, 40 Acad. Glushkova  avenue,  Kiev, Ukraine.
E-mail: yugo.ost@gmail.com }

\begin{abstract}

Modern progress in artificial intelligence permits to realize algorithms of adaptation for critical  events (in addition to ERP).
A production emergence, an appearance of new competitive goods, a major change in financial state of partners,
a radical change in exchange rate, a change in custom and tax legislation, a political and energy crisis, an ecocatastrophe can  lead up to a decrease of profit  
or bankruptcy of enterprise. Therefore it is necessary to assess a probability of threat and to take preventive actions.  
If a critical  event took place, one must estimate restoration expenses  and possible consequences as well as to prepare appropriate propositions.
This is provided using modern methods of diagnostics, prediction, and decision making as well as an inference engine and semantic analysis.
Mathematical methods in use are called in algorithms of adaptation automatically.
Because the enterprise is a complex system,  to overcome  complexity of control it is necessary to apply semantic representations. 
Such representations are formed from descriptions of events, facts, persons, organizations, goods, operations, scripts on a natural language.
Semantic representations permit as well to formulate actual problems and  to find ways to resolve these problems.

\end{abstract}

\end{frontmatter}

\section{Introduction}

The given paper is devoted to control problems connected  with the reaction to critical events that disrupt normal work of enterprise.
In living enterprise control systems (ERP) this question, as a rule, is not taken into consideration in full measure.
At the same time, the adaptation of control system to critical events plays a leading role in provision of financial stability and, 
consequently, survival under conditions of strong competition.
Because the enterprise is a complex system, we consider briefly the problem of adaptation and decision making in complex systems.
A complex system satisfies roughly the such conditions:

\begin{itemize}

\item  in the first place, the system consists of the large number of heterogeneous  components joined with a general purpose;
\item  in the second place, many components have not mathematical descriptions;
\item  in the third place, the system is exposed to internal and external actions.
                                                                                  
\end{itemize}

 To control an enterprise as a complex system one must realize the next functions:\\[10pt] 
\hspace*{20pt} 1. {\it Gathering information}. The input of data about a state of system is implemented using the measurement of factors as well as the description of facts (on formalized and natural language).
Information is usually saved in a database.\\
\hspace*{20pt} 2. {\it Planning}. Planning can be strategic and day-to-day. Strategical planning is based on forming decision variants and their estimation. Day-to-day planning consists in planning production, logistic, and marketing operations. \\ 
\hspace*{20pt} 3. {\it Executing plans}. To realize strategic plans the regular control of financial and production results is executed to change these plans as the need arose. 
Day-to-day planning is based on the correction of deviations from normal course of production, logistic, and marketing process.  \\
\hspace*{20pt} 4. {\it Reaction to critical events}. Adaptation algorithms estimate threats, their causes and consequences as well as form propositions to resolve available problems.\\[10pt] 
 
By this means, adaptive enterprise control includes three important phases. The first phase is the estimation of enterprise state: belonging to a certain class ({\it diagnostics}).
The second phase is the {\it  prediction} of system state by means of behaviour simulation. 
The simulation is founded on the measurement of basic factors that influence the system behaviour. 
The third phase is {\it decision making} using the prediction and inference. The inference permits to compare different variants of decisions and scripts of event development.  

General  questions of complex system adaptation are considered in \cite {Princip}.
According to {\it the law of requisite variety}\cite {Ashby}, to live and achieve a success complexity and quickness of decision making must correspond to complexity and  quickness of changes in environment.
One of way to overcome complexity of control is a mathematical simulation. For example, a variety of prices and volumes of sales  can be decreased at the cost of forming dependence between a price and a volume.
By this means, to resolve more and more complex tasks one must build more and more complex control system using modern mathematical methods to describe regularities of environment.
Because a social system is described the most adequately by means of a natural language, the control of such system will be effectively only using the semantic representation\cite {Ostapov1,Ostapov2,Ostapov3}.

As applied to enterprises, the control problem for critical events (risk-management) is considered in \cite {Ackoff,Ansoff,Baldin}. Algorithms of processing for production critical events are given in \cite {Jampol}.
The software for record and processing of risks is designed with the company SAP \cite {SAP}. 
The general distinction of our technology is in the comprehensive approach to the analysis of critical events and forming propositions directed to the solution of available problems.
In the end, only the comprehensive approach provides successful work of enterprise. 

It should be pointed out that  one must use methods of informal task solution for adaptation algorithms  when such problems are described with a natural language \cite {Ostapov4}.
These algorithms are based on a semantic analysis and inference. The inference uses operations and scripts from a {\it knowledge base}.
An operation describes a sequence of actions to realize a certain purpose. A script presents actions of persons and organizations after a certain event. These actions are based on  branch  experience.
The description of critical event is inputted to a control system on a natural language.

\section{Algorithms of diagnostics, prediction and decision making}

A complex system is characterized with its state at the moment. This state is described with a vector $(x_{1},...,x_{n})$.
Components of vector are measured with the help of appropriate scales\cite {Malhot}.

Consider briefly some effective algorithms of diagnostics, prediction, and decision making for adaptive enterprise control. 

\subsection{Algorithms of diagnostics}

To solve diagnostics tasks {\it methods of pattern recognition} are used to establish belonging of  object state to a certain class.
A procedure that permits to assign a given state of object to a certain  class by means of the analysis of characteristics is referred to as an {\it algorithm of recognition}.
Values of characteristics are established according to the next rules:\\[10pt] 
\hspace*{20pt} 1. If there is a given characteristic for object under investigation, then the value of such characteristic is 1 .\\
\hspace*{20pt} 2. If this characteristic absents  for object, then the value of such characteristic is -1 .\\ 
\hspace*{20pt} 3. If there is not any information concerning this characteristic, then the value of such characteristic is 0.\\[10pt]

When the state of complex system is described with the set of such characteristics, this set will be referred to as a  {\it logic vector} $(x_{1},...,x_{n})$.
 
A {\it decision function} is a mathematical expression from characteristics to solve the recognition problem.
Parameters of this function are determined on the base of experience data using a {\it learning algorithm}.

We have restricted ourselves to the next methods:

\begin{itemize}

\item  {\it the method of separating plane;
\item  the method of separating surface;
\item  the method of frequencies;
\item  the method of potential functions}.
                                                                                  
\end{itemize}

Other approaches are considered in \cite {Luger,Russel}.

\subsubsection{The method of separating plane}

{\it The method of separating plane} is based on {\it the least-squares method} to find parameters of the decision function $\Phi$:
\begin{equation}
   \Phi = a_{0} + a_{1} \,  x_{1} + ...+ a_{n} \,  x_{n} .                                                           
\end{equation}

Let us explain what a separating plane, in fact, means. Consider a hypothesis concerning belonging of object state to a certain class.
The collection of points  $(x_{1},...,x_{n})$ that correspond to a given hypothesis forms a set in vector space of dimension $n$.
This set is referred to as an {\it image} of the given class.
One of approaches for the recognition problem is in forming a plane to separate  images of classes in space. 
To solve the recognition problem it is necessary to check to which space part an object state under investigation will be assigned.

The equation of separating  plane is $\Phi = 0$.
If the inequality $\Phi > \varepsilon$  is true, then the given object corresponds to the first hypothesis (concerning belonging to the first class)
\footnote{Hereafter $\varepsilon$ is a sufficiently small positive number.}. 
If the inequality $\Phi < - \varepsilon$  is true, then the given object corresponds to the second hypothesis (concerning belonging to the second class). 

{\it The method of separating surface} is characterized with the application of polynomial instead of a linear function. 
These geometrical methods were considered in \cite {Ivah}.

\subsubsection{The method of frequencies}

{\it The method of frequencies} is based on the linear decision function for the hypothesis with a number $j$:
\begin{equation}
   \Phi_{j} =   a_{1}^{j} \,  x_{1} + ...+ a_{n}^{j} \,  x_{n} ,  \, a_{i}^{j} = \ln{\frac{p_{i}^{j}}{ 1-p_{i}^{j}}} ,                                                        
\end{equation}
where $p_{i}^{j}$ is the frequency of appearance for characteristic $x_{i}$ in experience data for this hypothesis.

As a hypothesis that corresponds to the state of system,  we take the hypothesis with maximal $\Phi_{j}$ for $j=1,..,m$.
The decision function is calculated directly by means of available data. This method was proposed in \cite {Lerner}.

\subsubsection{The method of potential functions }

{\it The method of potential functions} is the most strong from methods under consideration as 
one can separate {\it  at will located images} in vector space of dimension $n$.
The distinctive feature of method is:

\begin{itemize}

\item  direct building a decision function  using a table of experience data;
\item  the application for several classes.

\end{itemize}

The method of potential functions is based on the next decision function:
\begin{equation}
   \Phi = \sum_{i=1} ^ {m} \gamma_{i}\, \varphi (x, x^{i}) ,                                                               
\end{equation}
where $x$ is a logical vector describing a given state of system, $x^{i}$ is a logical vector  from a table of experience data, 
$\gamma_{i}=1$ if the vector $x^{i}$ belongs to the first image, $\gamma_{i}=-1$ if the vector $x^{i}$ belongs to the second image, 
 $\varphi (x, a)$ is a potential function for a vector $x=(x_{1},...,x_{n})$ and a vector $a=(a_{1},...,a_{n})$.
The potential function $\varphi (x, a)$ is determined by the expressions:
\begin{equation}
   \varphi (x, a)= \frac{1}{\rho^{2}(x,a)}, \,\rho^{2}(x,a)\geq\varepsilon,                                                                
\end{equation} 

\begin{equation}
   \varphi (x, a)= \frac{1}{\varepsilon}, \,\rho^{2}(x,a)\leq\varepsilon,                                                                
\end{equation} 
where
\begin{equation}
    \rho^{2}(x,a)= \sum_{j=1}^{n}(x_{j} - a_{j})^{2} .                                                             
\end{equation}

If the inequality $\Phi > \varepsilon$  is true, then the given system corresponds to the first hypothesis (concerning belonging to the first class). 
If the inequality $\Phi < - \varepsilon$  is true, then the given system corresponds to the second hypothesis (concerning belonging to the second class).  
This method was considered in \cite {Encycl}.

\subsection{Algorithms of prediction}

Preparatory to decision making, it is necessary to execute the predication  of basic activities.
To do this, the {\it linear regression analysis} is frequently used. The next expression connects the factor $y$ with characteristics  $x_{1},...,x_{n}$:
\begin{equation}
   y =   a_{1} \,  x_{1} + ...+ a_{n} \,  x_{n} .                                                         
\end{equation}

This model is true only for small changes of characteristics. The parameters $a_{i}$ are determined using {\it the least-squares method}.
In the general case, a nonlinear model is more adequate:
\begin{equation}
   y =   f(x_{1}, ..., x_{n}) .                                                         
\end{equation}
 
If the right part (8) is polynomial, then we use  the {\it nonlinear regression analysis}\cite {Eddous}.
Reliability of predication is investigated with the methods of {\it mathematical statistics}.
High reliability of prediction is provided, first, with the sufficient number of data, second, with the correlation between $y$ and $x_{i}$, third,
with the small dispersion of $y$ connected with regression\cite {Malhot}.
  
If the expression (8) can not be presented with polynomial, then other algorithms are used.
Consider the application of potential function method. Divide the interval for changes of $y$ into 100 segments.
Belonging of $y$ to some segment is the hypothesis about accordance of a certain class. In this case the accuracy of prediction is 0.5 percent.

It can use as well a {\it dynamical model} that permits to predict the behaviour of system at several future moments:
\begin{equation}
   y(t+1) = \sum_{i=0}^{m} a_{i}\, y(t-i) + \sum_{j=1}^{n} b_{j}\, v_{j}(t) ,                                                                
\end{equation} 
where $t$ is  discrete time (positive integer). For an enterprise it can use, for example, such characteristics: $y$ is a profit,
$v_{1}$ is the cost of energy supply, $v_{2}$ is the sale volume of general competitor, $v_{3}$ is rate of exchange, etc.
The parameters $a_{i}$ and $b_{j}$ are determined with the experience data concerning the enterprise activity using {\it the least-squares method}. 

Other prediction methods are considered in the technology {\it Data Mining}\cite {Kantar}.
To take account of new factors one should be used an inference (see further).

\subsection{Algorithms of decision making}

Decision making is a {\it choice} from several variants. Each variant has certain properties, and the problem is in the {\it optimal choice}. 
The final result is a decision -- the prescription for action.

The choice of decision is connected with the analysis of external factors exerting influence on choice.
For example, for an enterprise such factors will be the percent of inflation, characteristics of demand, size of taxes, etc.
The state of environment is described with the set of alternative situations.

By this means, to make a decision it is necessary to find the optimal variant based on the criterion of choice and taking account of external factors.
To estimate variants a {\it function of preference} is used. This function is calculated for each variant and situation.

\subsubsection{ The table   "Decision-Situation"}
 
The table  {\it Decision-Situation} is applied to present values of the {\it function of preference}.
We must:
\begin{itemize}

\item  to indicate the list of decision variants;
\item  to point the list of situations under investigation;
\item  to calculate the function of preference $V_{ij}$ for each variant $i$ and a certain situation $j$.
                                                                                  
\end{itemize}

The {\it pessimistic choice} is based on the {\it principle of minimax}. For each variant $i$ we suppose the worse case when the function $V_{ij}$
is minimal for $j=1,...,m$ ($j$ is a number of situation). Then we find the variant $k$  when the function $V_{ij}$ is maximal:
\begin{equation}
   w_{k} =   max (min V_{ij}), i=1,...,n .                                                         
\end{equation}

The {\it optimistic choice} is based on the {\it principle of maximax}. For each variant $i$ we suppose the best case when the function $V_{ij}$
is maximal for $j=1,...,m$:
\begin{equation}
   w_{k} =   max (max V_{ij}), i=1,...,n .                                                         
\end{equation}

If the probability of each situation $j$ is $P_{j}$, then the {\it pragmatic choice} is realized as follows:

\begin{equation}
   w_{k} =   max (\sum_{j=1}^{m} P_{j}\, V_{ij}), i=1,...,n .                                                         
\end{equation}

\subsubsection{ Assessment of probability for critical events}

In order in time to take preventive actions it is necessary to estimate a probability of critical event.
Such assessment can be executed with the methods of {\it probability theory}. Consider, for example, the {\it Bayesian formula}\cite {Luger}:
\begin{equation}
   P(H_{i}/E) =   \frac{P(H_{i})\,P(E/H_{i})}{ \sum_{j=1}^{n} P(H_{j})\,P(E/H_{j})} .                                                        
\end{equation}
where $H_{i}$ is the hypothesis  with a number  $i$ for a certain critical event, $E$ is the event that serves as a preventive signal.

\subsection{Inference engine}

An inference engine realizes algorithms of adaptation. This engine implements two sorts of inference:
\begin{itemize}

\item  direct -- on the basis of the sentential calculus;
\item  indirect -- using the predicate calculus.

\end{itemize}

\subsubsection{Direct inference}

The direct inference is realized by means of rules:
\begin{equation}
    A_{1}^{j} \wedge A_{2}^{j} \wedge ...  \wedge A_{n_{j}}^{j} \to B_{j}, j=1,...,m ,                                                         
\end{equation}
where $B_{j}$ and $A_{i}^{j}$ are logic statements.

Such rules can be presented graphically with a decision tree\cite {Kantar}. 
A tree node corresponds to  a logic statement. Such approach permits to realize an inference when an user answers "Yes" or "No" for questions
to be formed by a program. 

\subsubsection{Indirect inference}

The indirect inference is based  on rules\footnote{The method of resolution for Horn's sentences is used\cite {Luger}.}:
\begin{equation}
  \psi_{j} \gets  \varphi_{1}^{j}, \varphi_{2}^{j}, ... , \varphi_{n_{j}}^{j}, j=1,...,m ,                                                         
\end{equation}
where $\psi_{j}$ and $\varphi_{i}^{j}$ are predicates.

To explain the inference algorithm a goal tree is built. A tree node corresponds to a predicate (an executed goal).
To present each rule (15) nodes for predicates $\psi_{j}$ and $\varphi_{i}^{j}$ are connected by means of edges. 
The algorithm of inference consists in sequential check of all predicates for tree traversal top-down and from left to right.
Because a predicate is true when the appropriate goal is executed, the algorithm is in sequential check of execution for all goals of tree 
\footnote{This algorithm corresponds to the absence of  alternatives.
If there are alternatives, the algorithm gets complicated by the application of return when some goal is not executed.}.

\subsubsection{Predicates of inference engine}

An inference engine is realized by means of the program package {\it Advanced Problem Solver}.
To build algorithms of adaptive control the next predicates are needed:
\begin{itemize}

\item to select information from a database;
\item to calculate arithmetic expressions;
\item to compare intermediate results;
\item to process dynamical arrays;
\item to organize dialogue with an user;
\item to analyze data (pattern recognition, modelling).
                                                                                  
\end{itemize}

\section{Algorithms of adaptation for critical  production events}

Algorithms of adaptation are required for the next types of events:

\begin{itemize}

\item  extraordinary production situations having serious consequences (conflagrations, explosions, destructions of aggregates and constructions, 
chemical and radioactive contamination);
\item damages of technological equipment;
\item failures of infrastructure (transport, connection, control, energy supply).
                                                                                  
\end{itemize}

It should be pointed out that one must apply semantic methods to analyze descriptions of critical events on a natural language\cite {Ostapov1,Ostapov2,Ostapov3,Ostapov4}. 
Algorithms of inference used in such methods are based on branch experience and economical laws. 

We have restricted ourselves only to the consideration of extraordinary production situations as for other types of events algorithms are formed analogously.

In the case of extraordinary production situations, the basic goal is divided into two alternatives
\footnote {Hereafter to describe algorithms of adaptation we use rules of indirect inference. Goals under consideration are presented by predicates.}:

\begin{equation}
  \Phi \gets  \Phi_{1}, \Phi_{2} ,\quad \Phi \gets  \Phi_{3}, \Phi_{4} ,                                                             
\end{equation}
where $\Phi_{1}$ is  the processing of signals and the estimation of critical event probability,
$\Phi_{2}$ is the reaction for a certain level of threat, $\Phi_{3}$ is the semantic analysis of event description,
$\Phi_{4}$ is the reaction to the event.

The probability of emergency is assessed with the event tree\cite {Eddous} or the Bayesian formula (13). The reaction to a real threat
includes two subgoals:
\begin{equation}
  \Phi_{2} \gets  \Psi_{1}, \Psi_{2} ,                                                               
\end{equation}
where $\Psi_{1}$ is  the assessment of consequences, $\Psi_{2}$ is the preparation of propositions concerning preventive measures.
The assessment of consequences and the preparation of propositions are based on branch experience for analogous  damages.

The reaction to the available critical event has five subgoals:
\begin{equation}
  \Phi_{4} \gets  \Omega_{1}, \Omega_{2}, \Omega_{3}, \Omega_{4},\Omega_{5} ,                                                               
\end{equation}
where $\Omega_{1}$ is the description and assessment of measures after damages, $\Omega_{2}$ is  the analysis of emergency causes, 
$\Omega_{3}$ is the assessment of consequences, $\Omega_{4}$ is the preparation of propositions for the correction of plans (sales and production),
$\Omega_{5}$ is the preparation of propositions concerning the reliability of aggregates and constructions.

The assessment of measures for restoration proceeds from the character of available damages. Logic of reconstruction is based on a certain sequence of actions for the solution of problems under consideration.
The generation of plans is in the fragmentation of the given problem into tasks and the analysis of consequences.
The analysis of emergency causes proceeds from possible versions and check of conditions when these versions can be realized.

The estimation of consequences  is based on  the analysis of negative profit as a result of available destructions.
The preparation of propositions for the correction of plans proceeds from the analysis of consequences.
Forming propositions for reliability improvement is founded on branch experience.

By way of illustration, we consider a damage at a metallurgy enterprise:

{\it At the third blast furnace an explosion took place during the preparation to stop for the planned overhaul. 
The explosion of the gas mixture in the cupola of furnace duster caused the failure of massive metal construction to the gas pipeline, waterway, electrical cables. 
Gas ignited, the fire arose. The tank assigned for oil discharge and  heating main from the second pump station took fire. 
Because of waterway damage the electric substation was flooded so that it  turned off. The pump station of the first blast furnace was flooded as well and 
this station  was disabled. Because of the absence of water cooling all tuyeres of the first and fourth blast furnaces were damaged.  These furnaces were as well stopped.}

 The analysis and assessment of measures are based on experience of metallurgy enterprises concerning the breakdown elimination. This information must be contented in the database.
  
The next measures are proposed after the fire is extinguished:
\begin{itemize}

\item  to pump out water from  constructions;
\item  to restore the waterway;
\item  to restore and start the electric substation;
\item  to restore and start the pump station of the first blast furnace;
\item  to restore the tank assigned for oil discharge and the heating main from the second pump station;
\item  to restore all tuyeres of the first and fourth blast furnaces;
\item  to start   the first and fourth blast furnaces; 
\item  to restore the gas pipeline,  electrical cables of the third blast furnace.
                                                                   
\end{itemize}

For each measure, the sheet of expenses is formed to take decisions and get entitlement payments if the property of enterprise is insured against damages.

The analysis of emergency causes must be founded on the model that describes the physical-chemical process at the duster.
A received conclusion must take into account results of parameter measurement for blast-furnace gas at the moment of emergency. 

Propositions for the correction of production plans use information about the decrease of the output volume for cast iron.
The  volume of output is calculated based on terms of equipment restoration.

The assessment of consequences includes the next factors:\\[10pt]
\hspace*{20pt} 1. The change of sale volume for output goods: slabs, pipes, rolled stock.\\
\hspace*{20pt} 2. The penalty sanctions because of delivery delay;\\ 
\hspace*{20pt} 3. The increase of account payable if a credit is required.\\[10pt]
 
Forming propositions concerning reliability improvement follows from the analysis of causes for the given emergency.
For example, taking into account  available experience, it can  change  the construction of duster.

\section{Algorithms of adaptation for critical  events on the market}

Algorithms of adaptation are realized for the next types of events:
\begin{itemize}

\item an appearance of new competitive goods;
\item opening a new segment of market;
\item a major change in the financial state of partners and competitors.
                                                                                  
\end{itemize}

Algorithms of adaptation for critical  events on the market apply semantic methods for the control problem solution\cite {Ostapov1,Ostapov2,Ostapov3,Ostapov4}.
Information about an appearance of new competitive goods, opening a new segment of market and the change in the financial state of partners and competitors must be known beforehand. 
This is provided with the help of market monitoring.

In the case of appearance of new competitive goods, the basic goal includes four subgoals:
\begin{equation}
  \Phi \gets  \Phi_{1}, \Phi_{2}, \Phi_{3}, \Phi_{4} ,                                                             
\end{equation} 
where $\Phi_{1}$ is the assessment of consumer value for competitive goods, $\Phi_{2}$ is the assessment of influence on own sales, 
$\Phi_{3}$ is the preparation of information for the correction of plans, $\Phi_{4}$ is the preparation of  propositions (as the need arose) for the application of new technology.

The assessment of consumer value is executed by means of the regression model\cite{Malhot}. 
The estimation of influence on own sales is analyzed as well using the regression analysis (see above).
Forming information for the correction of plans is based on results concerning influence on sales. 
Propositions for the application of new technology follow from the assessment of consumer value for competitive goods.

In the case of opening a new segment of market, the basic goal has three subgoals: 
\begin{equation}
  \Phi \gets  \Phi_{1}, \Phi_{2}, \Phi_{3},                                                             
\end{equation} 
where $\Phi_{1}$ is the analysis of  new segment using methods in \cite{Ansoff}, $\Phi_{2}$ is the assessment of the sale volume by means of the regression analysis(see above), 
$\Phi_{3}$ is the preparation of information for the correction of plans.

Under the change in the financial state of partners and competitors, the basic goal is divided into four subgoals:
\begin{equation}
  \Phi \gets  \Phi_{1}, \Phi_{2}, \Phi_{3}, \Phi_{4}                                                              
\end{equation} 
where $\Phi_{1}$ is the assessment of  financial state (partners and competitors), $\Phi_{2}$ is the assessment of   consequences, 
$\Phi_{3}$ is the preparation of information for the correction of plans, $\Phi_{4}$ is the preparation of other propositions.

The assessment of  financial state is executed by means of {\it Altman's formula}\cite{Baldin} or methods of diagnostics(see above).
The estimation of consequences is based on expert judgement (for example, using the decision tree). 
Information for the correction of plans is founded on results of the consequence assessment. 
 
\section{Algorithms of adaptation for critical  events in regions}

Algorithms of adaptation are used for the next types of events:
\begin{itemize}

\item a radical change in exchange rate;
\item a change of customs fees;
\item a change in tax legislation;
\item a political crisis;
\item a  fuel and energy crisis;
\item an ecocatastrophe threatening  the enterprise, its partners and competitors. 
\end{itemize}

Algorithms of adaptation for critical  events in a region are based on semantic methods of control problem solution\cite{Ostapov1,Ostapov2,Ostapov3,Ostapov4}.

For a possible radical change in exchange rate, the basic goal includes two subgoals:
\begin{equation}
  \Phi \gets  \Phi_{1}, \Phi_{2} ,                                                            
\end{equation} 
where $\Phi_{1}$ is the prediction of radical change, $\Phi_{2}$ is the assessment of consequences.

The prediction of radical change in exchange rate  can be executed by means of the dynamical model (9).
The assessment of consequences consists in the determination of unprofitability (or profitability) for output goods after such radical change.
To do this,  it is necessary to implement expense accounting based on the cost of components, materials, labour, energy supply,  and logistic services\cite{Drury}.
To estimate a demand the regression analysis should be applied (see above). 

Information concerning changes in custom and tax legislation or the cost of energy supply is usually known beforehand.
Algorithms of adaptation must assess consequences of such changes -- to find unprofitable goods.

This is crucial as well to execute monitoring of political events in the region and to estimate consequences of major changes for enterprise activity.
The prediction of changes in a  political situation can be made using methods of political science\cite{Manheim}.
The estimation of consequences is in the determination of possible decrease for sales, an appearance of logistic problem, etc. 

An ecocatastrophe demands algorithms that are similar to algorithms for a production emergency.

\section{Conclusion}

Modern progress in artificial intelligence permits to realize algorithms of adaptation for critical  events (in addition to ERP).
This is provided using:
\begin{itemize}

\item   modern methods of diagnostics, prediction and decision making;
\item   an inference engine;
\item   a semantic analysis. 

\end{itemize}

It should be emphasized that mathematical methods in use are called automatically.
The proposed technology  provides economical stability of enterprise under  conditions of strong competition.

Because the enterprise is a complex system consisting of persons and hardware, to overcome  complexity of control it is necessary to apply semantic representations for descriptions on a natural language.
Such representations permit to formulate actual problems and  to find methods to resolve these problems.

\end{document}